\documentclass[11pt]{article}
\usepackage[a4paper,margin=1in]{geometry}
\usepackage[T1]{fontenc}
\usepackage[utf8]{inputenc}
\usepackage{lmodern}
\usepackage{microtype}
\usepackage{booktabs}
\usepackage{array}
\usepackage{longtable}
\usepackage{enumitem}
\usepackage[hidelinks]{hyperref}
\usepackage{xcolor}
\usepackage{titlesec}
\usepackage{abstract}
\setlist{nosep,leftmargin=*}
\titleformat{\section}{\large\bfseries}{\thesection.}{0.5em}{}
\titleformat{\subsection}{\normalsize\bfseries}{\thesubsection.}{0.5em}{}

\newcolumntype{L}[1]{>{\raggedright\arraybackslash}p{#1}}

\title{From Battlefield to Boardroom: Strategic Red Teaming as an Epistemic Governance Instrument in the Age of AI}
\author{Jeroen Janssen\\Apparens\\\texttt{office@apparens.nl}}
\date{arXiv technical report edition, July 2026}

\begin{document}
\maketitle

\begin{abstract}
Strategic planning for AI adoption commonly treats approval as sufficient evidence that a proposed initiative is sound. This technical report formalizes strategic red teaming as a board-level assumption stress-testing model for AI governance. The model is designed to test the propositions that must be true for a material AI strategy to be defensible, rather than to test model outputs, security controls, or compliance artefacts alone. The report preserves the original title of the working paper but reframes the contribution as a bounded governance design artefact: a six-component process for mission-alignment testing, assumption mapping, dependency stress testing, economic-fragility testing, regulatory-exposure simulation, and accountability-boundary testing.

The argument is that AI integration changes the evidentiary burden of strategic approval along five exposure dimensions: operational leverage, transparency reduction, dependency concentration, regulatory liability, and accountability-boundary shift. Under these conditions, conventional risk registers and management-owned assurance processes are insufficient because they document known risks within an accepted plan rather than adversarially testing the assumptions that make the plan appear acceptable. The proposed design specifies inputs, outputs, evidence grades, independence requirements, reporting lines, and board decision records. It does not claim legal sufficiency, empirical validation, regulatory endorsement, or universal necessity. It claims a testable institutional design for producing board-quality evidence about the strength of strategic assumptions in AI-embedded organizations.
\end{abstract}


\section{Introduction}
AI adoption strategies often move from executive advocacy to board approval without an adversarial test of the assumptions on which the strategy depends. Standard risk governance usually asks whether identified risks have controls. It rarely asks whether the strategic propositions that justified the initiative were ever true enough to rely on. This report treats that gap as an epistemic governance problem.

The central question is: under what conditions does AI integration transform strategic red teaming from a useful governance instrument into a board-level assumption stress-testing requirement? The paper does not assert that every AI deployment requires a full strategic red-team engagement. It defines the conditions under which the evidentiary burden becomes material: consequential AI use, significant capital commitment, dependency on external AI providers, high-risk regulatory exposure, opacity in decision pathways, or an accountability structure that cannot be reconstructed after harm occurs.

The report contributes a design artefact: a formal model for strategic red teaming as independent, evidence-graded assumption stress testing. The object of test is not a model, application, policy, or control framework. The object is the cognitive architecture of a strategic decision: the propositions that must hold for the strategy to be defensible.

\section{Positioning and claim boundary}
This is a technical report on institutional design. It is not an empirical study, legal opinion, or regulator-endorsed framework. Its claims are bounded as follows.

\begin{itemize}
  \item \textbf{Design claim.} Strategic red teaming can be specified as a repeatable governance process with defined inputs, independence rules, evidence grades, outputs, and reporting lines.
  \item \textbf{Necessity claim.} A full engagement is warranted only when AI integration creates material uncertainty, opacity, dependency concentration, regulatory exposure, or accountability ambiguity.
  \item \textbf{Non-sufficiency claim.} A red-team finding does not establish compliance, safety, or legality. It establishes the evidence quality of assumptions and the conditions under which those assumptions may fail.
  \item \textbf{Validation status.} The artefact requires empirical validation in live board-governance settings and comparative testing against pre-mortem, audit, assurance, and risk-review methods.
\end{itemize}

\section{The governance problem}
Three recurrent features of strategic approval create the need for adversarial assumption testing.

First, approval processes are usually advocacy-driven. Strategy owners present the case for action, often supported by business cases, vendor claims, market narratives, and quantified benefits. These materials may be internally consistent while still resting on weak assumptions.

Second, AI changes the relation between decision, execution, and evidence. A strategic commitment to AI is not merely a technology purchase. It can delegate decisions to systems whose behaviour is difficult to explain, embed vendor dependencies in core workflows, alter accountability chains, and create regulatory obligations that outlive the original business case.

Third, conventional risk registers usually operate inside the accepted strategy. They ask which risks threaten delivery. Strategic red teaming asks an earlier question: which assumptions must be true for the strategy to deserve approval at all?

\section{Red teaming modalities}
Strategic red teaming should be distinguished from adjacent practices. The differences matter because each modality tests a different object and answers a different governance question.

\begin{table}[h]
\centering
\small
\begin{tabular}{L{0.23\textwidth} L{0.28\textwidth} L{0.31\textwidth} L{0.12\textwidth}}
\toprule
\textbf{Modality} & \textbf{Object of test} & \textbf{Governing question} & \textbf{Level} \\
\midrule
Penetration testing & Infrastructure, applications, networks & Can an adversary breach or misuse the technical perimeter? & CISO / IT security \\
AI red teaming & Model outputs, robustness, alignment, misuse behaviour & Can the model be induced to produce unsafe or misaligned outputs? & AI safety / technical teams \\
Governance red teaming & Oversight structures, controls, accountability mechanisms & Does the governance architecture function under adversarial conditions? & Risk / board committee \\
Strategic assumption stress testing & Strategic intent and enabling assumptions & Which assumptions must be true for this AI strategy to succeed, and what is the evidence quality for each? & Board / executive \\
\bottomrule
\end{tabular}
\caption{Red-teaming modalities by object, question, and governance level.}
\end{table}

\section{AI as an exposure multiplier}
The model treats AI as a strategic exposure multiplier along five dimensions.

\begin{enumerate}
  \item \textbf{Operational leverage.} AI can scale a decision rule, workflow, recommendation, or classification across many transactions. Weak assumptions therefore propagate faster and further than in human-only processes.
  \item \textbf{Transparency reduction.} AI can replace auditable human reasoning with probabilistic or opaque computational behaviour. The evidentiary burden shifts from explaining a human decision to reconstructing a socio-technical decision chain.
  \item \textbf{Dependency concentration.} Core workflow capability may become dependent on model providers, cloud services, data suppliers, integrators, or proprietary tooling. Vendor continuity becomes strategic continuity.
  \item \textbf{Regulatory liability.} Deployer obligations under AI regulation and sectoral law may attach to the organization even where the model is externally supplied or poorly understood.
  \item \textbf{Accountability-boundary shift.} Responsibility can move across developer, deployer, user, board, vendor, and affected person without being explicitly reassigned before deployment.
\end{enumerate}

These dimensions do not prove that a strategy will fail. They indicate where ordinary risk governance may under-specify the assumptions on which strategic approval depends.

\section{The six-component model}
The proposed artefact contains six components. Each component has required inputs, a process design, a primary output, and a governance position.

\begin{longtable}{L{0.20\textwidth} L{0.24\textwidth} L{0.28\textwidth} L{0.20\textwidth}}
\toprule
\textbf{Component} & \textbf{Required inputs} & \textbf{Process design} & \textbf{Primary output} \\
\midrule
\endfirsthead
\toprule
\textbf{Component} & \textbf{Required inputs} & \textbf{Process design} & \textbf{Primary output} \\
\midrule
\endhead
Mission-alignment testing & Strategy documents; AI deployment rationale; expected outcomes & Test whether AI capability serves mission or drives it; identify legitimacy-driven adoption & Mission-coherence assessment with evidence grade \\
Assumption mapping & Business case; forecasts; vendor claims; architecture choices & Identify enabling assumptions; classify as documented, inferred, asserted, or contradicted & Assumption register with falsification conditions \\
Dependency stress testing & Vendor contracts; architecture maps; operational critical paths & Construct single-point and correlated-failure scenarios; test exit and recovery capacity & Dependency heat map and recovery evidence gaps \\
Economic-fragility testing & Financial projections; sensitivity ranges; capital commitments & Apply downside pressure to benefits, cost, adoption, and switching assumptions & Break-even and irreversibility analysis \\
Regulatory-exposure simulation & AI portfolio; use-case classification; obligations matrix & Test whether regulatory obligations are known, owned, evidenced, and timed & Compliance-gap and obligation-readiness register \\
Accountability-boundary testing & RACI/RASCI; decision logs; vendor roles; human oversight design & Trace who can explain, intervene, override, approve, and remediate & Accountability-boundary map with unresolved ownership findings \\
\bottomrule
\caption{Six-component model for strategic red teaming.}
\end{longtable}

\section{Evidence grading}
Strategic red teaming produces evidence-quality findings, not predictions. The model therefore uses a simple evidence scale.

\begin{table}[h]
\centering
\small
\begin{tabular}{L{0.20\textwidth} L{0.57\textwidth} L{0.15\textwidth}}
\toprule
\textbf{Grade} & \textbf{Definition} & \textbf{Board implication} \\
\midrule
Documented & Supported by current, traceable, decision-relevant evidence & Defensible unless contradicted \\
Tested & Supported by operational test, simulation, pilot, audit, or independent review & Stronger basis for approval \\
Inferred & Derived from analogous evidence or expert judgement, but not directly tested & Requires explicit uncertainty note \\
Asserted & Stated without sufficient evidence or testable support & Should not be treated as decision-grade \\
Contradicted & Evidence exists against the assumption or shows material inconsistency & Requires redesign, rejection, or explicit risk acceptance \\
Unknown & Evidence unavailable, inaccessible, or withheld & Treated as finding in itself \\
\bottomrule
\end{tabular}
\caption{Evidence-grade scale for strategic assumptions.}
\end{table}

\section{Independence and reporting architecture}
The model requires structural independence. Without it, red teaming can become a performative confirmation exercise. The minimum reporting architecture is:

\begin{enumerate}
  \item The board audit committee or equivalent independent committee commissions the engagement.
  \item The scope is recorded in board minutes and cannot be limited by the strategy owner.
  \item The red team has access to strategic documents, business cases, vendor agreements, architecture maps, risk registers, and compliance artefacts.
  \item Management may correct factual errors but may not delete, soften, or rewrite findings.
  \item Findings are presented directly to the board committee, with management response separated from the red-team report.
  \item The red team has no implementation advisory role or continuing commercial incentive tied to strategy continuation.
\end{enumerate}

\section{Board decision record}
A board decision following a strategic red-team engagement should record at least:

\begin{itemize}
  \item the strategy or AI commitment under review;
  \item the assumptions tested and their evidence grades;
  \item material assumptions graded asserted, contradicted, or unknown;
  \item unresolved regulatory, dependency, or accountability exposures;
  \item management response and factual corrections;
  \item board decision: approve, approve with conditions, defer, redesign, or reject;
  \item conditions for re-review and triggering events.
\end{itemize}

This record is not merely administrative. It is the durable link between strategic conviction, known uncertainty, and board accountability.

\section{Threats to validity}
\begin{itemize}
  \item \textbf{Conceptual status.} The model is a design artefact and has not yet been empirically tested across organizations.
  \item \textbf{Jurisdictional variation.} Fiduciary and regulatory implications vary by jurisdiction, sector, and corporate form.
  \item \textbf{Overuse risk.} The method is resource-intensive and should be reserved for material AI commitments, not routine low-risk automation.
  \item \textbf{Capture risk.} Red teams can become dependent on the organizations they review. Independence rules mitigate but do not eliminate this risk.
  \item \textbf{False precision.} Evidence grades must not be mistaken for probabilities of success or failure.
  \item \textbf{Validation need.} Comparative studies are needed against pre-mortem workshops, conventional risk reviews, internal audit, AI assurance, and external red-team engagements.
\end{itemize}

\section{Minimum validation protocol}
The artefact can be validated through four tracks.

\begin{enumerate}
  \item \textbf{Case reconstruction.} Apply the model retrospectively to failed or contested AI initiatives and test whether it would have surfaced the load-bearing assumptions before approval.
  \item \textbf{Board simulation.} Run controlled decision exercises in which boards review an AI strategy with and without the red-team artefacts, then compare assumption visibility and decision conditions.
  \item \textbf{Inter-rater reliability.} Have independent reviewers grade the same assumptions using the evidence scale and measure agreement.
  \item \textbf{Field pilot.} Apply the method to a live material AI decision and evaluate whether findings altered scope, controls, timing, vendor terms, or approval conditions.
\end{enumerate}

\section{Conclusion}
Strategic red teaming is not a substitute for technical AI red teaming, risk management, legal review, or audit. It is a missing prior step: independent, evidence-graded testing of the strategic assumptions that make an AI commitment appear defensible. The board-level question is not simply whether an AI system can be controlled after deployment. It is whether the organization has enough evidence to approve the strategic dependency it is creating.

The model proposed here converts that question into a repeatable governance artefact: a six-component process, an evidence-grade scale, an independence architecture, and a decision record. Its value is not certainty. Its value is disciplined adversarial clarity before organizational commitment becomes path dependency.

\appendix
\section{Strategic assumption register schema}
\begin{longtable}{L{0.28\textwidth} L{0.62\textwidth}}
\toprule
\textbf{Field} & \textbf{Purpose} \\
\midrule
\endfirsthead
\toprule
\textbf{Field} & \textbf{Purpose} \\
\midrule
\endhead
Assumption ID & Stable identifier for traceability. \\
Assumption statement & Proposition that must be true for the strategy to succeed. \\
Strategic dependency & Initiative, investment, workflow, vendor, or obligation dependent on the assumption. \\
Evidence grade & Documented, tested, inferred, asserted, contradicted, or unknown. \\
Evidence object & Source material supporting or refuting the assumption. \\
Falsification condition & Observable condition that would make the assumption false. \\
Exposure dimension & Operational leverage, transparency reduction, dependency concentration, regulatory liability, or accountability boundary. \\
Owner & Role accountable for evidence, not merely for delivery. \\
Board decision relevance & Why the assumption matters to approval. \\
\bottomrule
\end{longtable}

\section{Independence criteria}
A red-team engagement should be treated as structurally compromised if any of the following conditions apply: the team reports to the strategy owner; scope can be narrowed by management after commissioning; the team has a financial interest in implementation; findings can be edited by management before board presentation; access denials are not reportable as findings; or the same party performs both strategy advocacy and adversarial review.

\section{Board decision template}
\begin{enumerate}
  \item Strategy reviewed.
  \item Material AI exposure dimensions present.
  \item Assumptions tested.
  \item Assumptions graded asserted, unknown, or contradicted.
  \item Required evidence before approval.
  \item Required control or contractual changes.
  \item Decision and rationale.
  \item Conditions for re-review.
\end{enumerate}

\section*{Disclosure}
This report is an arXiv technical-report edition of an Apparens working paper. Apparens is the author's affiliation. The report is intended as a research artefact and does not constitute legal advice, regulatory endorsement, or a claim of empirical validation.


\begin{thebibliography}{99}
\bibitem{argyris1978} Argyris, C. and Sch\"on, D. A. (1978). \textit{Organizational Learning: A Theory of Action Perspective}. Addison-Wesley.
\bibitem{basel2018} Basel Committee on Banking Supervision. (2018). \textit{Stress testing principles}. Bank for International Settlements.
\bibitem{das2017} Das, S. (2017). \textit{The Devil's Advocate: Intelligence Analysis and Red Teaming}. Defence Studies literature.
\bibitem{dimaggio1983} DiMaggio, P. J. and Powell, W. W. (1983). The iron cage revisited: institutional isomorphism and collective rationality. \textit{American Sociological Review}, 48(2), 147--160.
\bibitem{eisenhardt1989} Eisenhardt, K. M. (1989). Agency theory: an assessment and review. \textit{Academy of Management Review}, 14(1), 57--74.
\bibitem{euai2024} European Union. (2024). Regulation laying down harmonised rules on artificial intelligence (Artificial Intelligence Act).
\bibitem{feffer2024} Feffer, M. et al. (2024). Red-teaming and evaluating generative AI systems. AI safety and evaluation literature.
\bibitem{ganguli2022} Ganguli, D. et al. (2022). Red teaming language models to reduce harms: methods, scaling behaviors, and lessons learned. \textit{arXiv preprint}.
\bibitem{goodfellow2015} Goodfellow, I., Shlens, J. and Szegedy, C. (2015). Explaining and harnessing adversarial examples. \textit{International Conference on Learning Representations}.
\bibitem{hood2001} Hood, C. and Rothstein, H. (2001). Risk regulation under pressure: problem solving or blame shifting? \textit{Administration \& Society}, 33(1), 21--53.
\bibitem{kahneman1993} Kahneman, D. and Lovallo, D. (1993). Timid choices and bold forecasts: a cognitive perspective on risk taking. \textit{Management Science}, 39(1), 17--31.
\bibitem{klein1993} Klein, G. (1993). A recognition-primed decision model of rapid decision making. In \textit{Decision Making in Action}.
\bibitem{knight1921} Knight, F. H. (1921). \textit{Risk, Uncertainty and Profit}. Houghton Mifflin.
\bibitem{lovallo2003} Lovallo, D. and Kahneman, D. (2003). Delusions of success: how optimism undermines executives' decisions. \textit{Harvard Business Review}.
\bibitem{march1958} March, J. G. and Simon, H. A. (1958). \textit{Organizations}. Wiley.
\bibitem{mintzberg1985} Mintzberg, H. and Waters, J. A. (1985). Of strategies, deliberate and emergent. \textit{Strategic Management Journal}, 6(3), 257--272.
\bibitem{nist2023} National Institute of Standards and Technology. (2023). \textit{Artificial Intelligence Risk Management Framework (AI RMF 1.0)}.
\bibitem{perrow1984} Perrow, C. (1984). \textit{Normal Accidents: Living with High-Risk Technologies}. Basic Books.
\bibitem{power2007} Power, M. (2007). \textit{Organized Uncertainty: Designing a World of Risk Management}. Oxford University Press.
\bibitem{schoemaker1995} Schoemaker, P. J. H. (1995). Scenario planning: a tool for strategic thinking. \textit{Sloan Management Review}, 36(2), 25--40.
\bibitem{simon1957} Simon, H. A. (1957). \textit{Models of Man}. Wiley.
\bibitem{tetlock2005} Tetlock, P. E. (2005). \textit{Expert Political Judgment}. Princeton University Press.
\bibitem{weick1995} Weick, K. E. (1995). \textit{Sensemaking in Organizations}. Sage.
\bibitem{zenko2015} Zenko, M. (2015). \textit{Red Team: How to Succeed by Thinking Like the Enemy}. Basic Books.
\end{thebibliography}
\end{document}